\documentclass[aps,prd,floatfix,superscriptaddress,showpacs,showkeys]{revtex4}
\usepackage{amssymb}
\usepackage{amsmath}
\usepackage{amsfonts}
\usepackage{epsfig}
\usepackage{graphicx}
\usepackage{tabularx}
\usepackage{latexsym}
\usepackage{graphics}
\usepackage{subfigure}
\usepackage{verbatim}
\usepackage{color}
\newcommand{\seq}{\begin{subequations}}
\newcommand{\sen}{\end{subequations}}
\newcommand{\eq}{\begin{eqnarray}}
\newcommand{\en}{\end{eqnarray}}
\newcommand{\ra}{\rangle}

\def\L2{\Lambda^2}

\def\eq{\begin{eqnarray}}
\def\en{\end{eqnarray}}

\def\L2{\Lambda^2}

\def\eq{\begin{eqnarray}}
\def\en{\end{eqnarray}}

\def\L{{\cal L}}
\def\M{{\cal M}}
\def\e{\epsilon}
\begin{document}

\title{\boldmath Selected strong decay modes of $Y(4260)$}

\author{Yubing Dong}
\affiliation{
Institute of High Energy Physics, Beijing 100049, P. R. China}
\affiliation{
Theoretical Physics Center for Science Facilities (TPCSF), CAS,
Beijing 100049, P. R. China}
\author{Amand  Faessler}
\affiliation{
Institut f\"ur Theoretische Physik,  Universit\"at T\"ubingen,\\
Kepler Center for Astro and Particle Physics, \\
Auf der Morgenstelle 14, D--72076 T\"ubingen, Germany}
\author{Thomas Gutsche}
\affiliation{
Institut f\"ur Theoretische Physik,  Universit\"at T\"ubingen,\\
Kepler Center for Astro and Particle Physics, \\
Auf der Morgenstelle 14, D--72076 T\"ubingen, Germany}
\author{Valery E. Lyubovitskij}
\affiliation{
Institut f\"ur Theoretische Physik,  Universit\"at T\"ubingen,\\
Kepler Center for Astro and Particle Physics, \\
Auf der Morgenstelle 14, D--72076 T\"ubingen, Germany}
\affiliation{Department of Physics, Tomsk State University,
634050 Tomsk, Russia}

\date{\today}

\begin{abstract}

In the present work the $Y(4260)$ resonance is considered
as a weakly bound state of a pseudoscalar $D$ and an axial $D_1$ charm meson.
We consider the two-body decay $Y(4260)\to Z_c(3900)^{\pm} + \pi^{\mp}$,
where $Z_c(3900)^\pm$ is treated as hadron molecule as well.
Moreover we compute the $Y(4260)$ decay modes $J/\psi \pi^+ \pi^-$,
recently observed by the BESIII Collaboration, and $\psi(2S) \pi^+ \pi^-$.
In the last process both the contact diagram with
$DD_1 \to \psi(nS) \pi^+ \pi^-$ and the resonance diagram with
$DD_1 \to Z_c(3900)^{\pm} + \pi^{\mp} \to \psi(nS) \pi^+ \pi^-$
are taken into account.

\end{abstract}



\maketitle

\section{Introduction}

The recent observation by three collaborations ---
BESIII~\cite{Ablikim:2013mio},
Belle~\cite{Liu:2013dau} and CLEO-c~\cite{Xiao:2013iha} --- of the
new resonance $Z_c(3900)^\pm$, and its neutral partner $Z_c(3900)^0$ by
CLEO-c~\cite{Xiao:2013iha}, stimulated
several studies of this state originating in different
theoretical structure assumptions. Since the observed state can be charged
and carries intrinsic charm the main proposition rests on an interpretation
either as a hadronic molecular or as a tetraquark
state ~\cite{Guo:2013sya,Dong:2013iqa}.

$Z_c(3900)^{\pm,0}$ states can also play a role in the decay dynamics of
the $Y(4260)$, which is also considered as a resonance outside the usual
charmonium spectrum.
This resonance was first observed by {\it BABAR}~\cite{Aubert:2005rm}
and later confirmed by the CLEO-c~\cite{Blusk:2006am}
and Belle~\cite{Abe:2006hf} Collaborations.
In the literature
different interpretations of the structure of this state were considered
(for an overview, see e.g. Ref.~\cite{Zhu:2005hp}):
molecular assignment --- $DD_1(2420)$ molecular
state~\cite{Zhu:2005hp,Ding:2008gr,Wang:2013cya},
$J/\psi K \bar K$ bound state~\cite{MartinezTorres:2009xb},
$\chi_c\rho$~\cite{Liu:2005ay} or
$\chi_c\omega$ molecular state~\cite{Yuan:2005dr},
charmonium interpretation~\cite{LlanesEstrada:2005hz},
tetraquark~\cite{Maiani:2005pe,Chiu:2005ey},
mixed charmonium-tetraquark state~\cite{Dias:2012ek},
nonresonant explanation of the $Y(4260)$ state (
interference of $\psi(4160)$ and $\psi(4415)$ charmonia
states)~\cite{Chen:2010nv}, hybrid $c\bar c g$~\cite{Kou:2005gt},
hybrid mesons (mixing of $c\bar c$ and
$c \bar c g$)~\cite{Iddir:2006sh}, and
baryonium $\Lambda_c^+\Lambda_c^-$
bound state~\cite{Qiao:2005av}.
Note that in the coupled-channel model proposed
in~\cite{vanBeveren:2006ih}, it was noticed that
there is no pole associated with the $Y(4260)$ state.
In addition, the production and decay of
$Y(4260)$ via $e^+e^-\to Y(4260)\to J/\psi\pi^+\pi^-$ was also studied
in Ref.~\cite{Close:2008xn}.
The study of this reaction chain can be an important check for the
inner structure of the intermediate state $Y(4260)$.

Based on the hadronic molecular scenario, in Ref.~\cite{Dong:2013iqa}
we considered the $Z_c(3900)$ and a possible $Z_c'(3950)$.
In a phenomenological Lagrangian
approach~\cite{Faessler:2007gv}-\cite{Dong:2012hc} we studied
the strong decay widths for $Z_c(3900)^{\pm}\to \psi(nS)+\pi^{\pm}$
or $h_c(mP)+\pi^{\pm}$.
To set up the bound state structure of the composite state we use the
compositeness condition~\cite{Weinberg:1962hj}-\cite{Anikin:1995cf}
which is the key ingredient of our approach.
In Refs.~\cite{Efimov:1993ei,Anikin:1995cf}
and~\cite{Faessler:2007gv}-\cite{Dong:2012hc} it was proved
that this condition is an important and successful quantum field theory tool
for the study of hadrons and exotic states as bound states of their
constituents. Here we adopt a hadronic molecular
structure for the $Y(4260)$ state where the composition is made up
of the pseudoscalar $D(1870)$ and the axial
$D_1(2400)$ charm mesons.

In this work we analyze the
strong two-body decay $Y(4260)\to Z_c(3900)^{\pm} + \pi^{\mp}$
and the three-body decays
$Y(4260)\to J/\psi+\pi^+\pi^-$ and
$Y(4260)\to \psi(2s)+\pi^+\pi^-$ by using the same
phenomenological Lagrangian approach developed in
Refs.~\cite{Faessler:2007gv}-\cite{Dong:2009tg}.
The states $Z_c(3900)$ and $Y(4260)$ are considered as molecular states.
In particular, we consider the $Z_c(3900)^\pm$ as hadronic molecules
as was previously discussed in ~Refs.~\cite{Guo:2013sya}. $Z_c(3900)^\pm$
together with the neutral partner $Z_c(3900)^0$ form the isospin triplet
with the spin and parity quantum numbers $J^P = 1^{+}$,
\eq
\label{Eq1}
|Z_c^I(3900)\ra &=& \frac{1}{2} \,
\Big| \bar{D}^{\ast} \, \tau^I \, D
+ \bar{D} \, \tau^I \, D^{\ast} \ra\,, \quad I = +, -, 0 \,.
\en
We also consider the $Y(4260)$ state as a isosinglet molecular state,
\eq
|Y(4260)\ra \ = \ \frac{1}{2} \,
\Big| \bar{D}_1 D + \bar{D} D_1 \Big\ra\,,
\en
with $J^P=1^{-}$.
Here $D = (D^+, D^0)$, $D^\ast = (D^{\ast\,+}, D^{\ast\,0})$,
$D_1 = (D_1^+, D_1^0)$ are the doublets of pseudoscalar,
vector and axial-vector $D$ mesons;
$\tau^{\pm,0}$ are the isospin
matrices defined in terms of the triplet of Pauli matrices $\tau^i$ as
\eq
\tau^\pm = \frac{1}{\sqrt{2}} (\tau^1 \mp i \tau^2)\,,
\quad \tau^0 = \tau^3 \,.
\en
This paper is organized as follows. In Sec.~II we briefly review the basic
ideas of our approach and show the effective Lagrangians for our calculations.
Then, in Sec.~III, we proceed to estimate the widths of the strong two-body
$Y(4260)\to Z_c(3900)^{\pm} + \pi^{\mp}$ and three-body
$Y(4260)\to \psi(nS)+\pi^+\pi^-$ decays. For the last case
we explicitly include the decay amplitude
$Y(4260)\to Z_c^{\pm}(3900)\pi^{\mp}\to \psi(nS)\pi^+\pi^-$.
In our analysis we approximately take into account the mass distribution
of the $Y(4260)$ state.
Finally we present our numerical results and compare to
recent limits set by experiment.

\section{Basic model ingredients}

Our approach to the possible hadron molecules $Z_c(3900)$ and
$Y(4260)$ is based on interaction Lagrangians describing the coupling
of the respective states to their constituents,
\eq\label{Lagr}
{\cal L}_{Z_c}(x)&=&\frac{g_{_{Z_c}}}{2}\, M_{Z_c} \,
\vec{Z_c}^\mu(x)\int d^4y \, \Phi_{Z_c}(y^2) \, \Big(
\bar D^{\ast}_{\mu}(x+y/2) \, \vec{\tau} \, D(x-y/2) \,+\,
\bar D (x-y/2) \, \vec{\tau} \, D^{\ast}_\mu(x+y/2) \Big)\,, \nonumber\\
{\cal L}_{Y}(x)&=&\frac{g_{_{Y}}}{2}\, M_{Y} \,
Y^{\mu}(x)\int d^4y \, \Phi_{Y}(y^2) \, \Big(
\bar{D}_{1\,\mu}(x+y/2) \, D(x-y/2) \, \,+\,
\bar{D}(x-y/2) \, D_{1\,\mu}(x+y/2)\Big )\,,
\en
where $\vec{Z_c} \, \vec{\tau} = Z_c^+ \tau^- + Z_c^- \tau^+ + Z_c^0 \tau^3$;
$y$ is a relative Jacobi coordinate, and $g_{_{Z_c}}$ and
$g_{_{Y}}$ are dimensionless coupling constants of $Z_c(3900)$ and
$Y(4260)$ to the molecular $\bar{D}D^{\ast}$ and $\bar{D}D_1$
components, respectively. Here, $\Phi_H(y^2)$
($H = Y, Z_c$) is the correlation function which describes the distributions of
the constituent mesons in the bound states. A basic requirement for the
choice of an explicit form of the correlation function $\Phi_H(y^2)$
is that its Fourier transform vanishes sufficiently
fast in the ultraviolet region of Euclidean space to render the Feynman
diagrams ultraviolet finite. For simplicity we adopt a Gaussian form for
the correlation functions. The Fourier transform of this vertex function
is given by
\eq
\label{corr_fun}
\tilde\Phi_H(p_E^2/\Lambda^2_H)
\doteq \exp( - p_E^2/\Lambda^2_H)\,,
\en
where $p_{E}$ is the Euclidean Jacobi momentum. $\Lambda_H$ stand for
the size parameters characterizing the distribution of the two constituent
mesons in the $Z_c(3900)$ and $Y(4260)$ systems.

From our previous analyses of strong two-body decays of $X, Y, Z$ meson
resonances interpreted as hadron molecules and of the $\Lambda_c(2940)$,
$\Sigma_c(2880)$ baryon states we deduced a value of about
$\Lambda \sim 1$~GeV~\cite{Dong:2009tg}. For a very loosely bound system
like the $X(3872)$, a size parameter of
$\Lambda \sim 0.5$~GeV~\cite{Dong:2009uf} is more suitable.
Once the size parameters and the masses of
the bound state systems are chosen, the respective coupling constants
$g_{_H}$ are determined by the compositeness
condition~\cite{Weinberg:1962hj,Efimov:1993ei,Anikin:1995cf, Dong:2009tg,
Faessler:2007gv}. It implies that the renormalization constant of the hadron
wave function is set equal to zero with
\eq\label{ZLc}
Z_H  = 1 - \Sigma_H^\prime(M_H^2) = 0 \,.
\en
Here, $\Sigma_H^\prime$ is the derivative of the transverse part of the
mass operator $\Sigma_H^{\mu\nu}$ of the molecular states
(see Fig.~1), which is defined as
\eq
\Sigma_H^{\mu\nu}(p) = g^{\mu\nu}_\perp \, \Sigma_H(p)
+ \frac{p^\mu p^\nu}{p^2}
\Sigma_H^L(p)\,, \quad
g^{\mu\nu}_\perp = g^{\mu\nu} - \frac{p^\mu p^\nu}{p^2} \,.
\en
The explicit expression for the coupling constant $g_H$ $(H = Y, Z_c)$
resulting from the compositeness condition is
\eq
g^{-2}_{H}&=& \frac{M_H^2}{\Lambda_H^2} \, \int\limits_0^\infty
\frac{d\alpha d\beta}{16\pi^2\Delta^2} \, R(\alpha,\beta) \,
\biggl( 1+\frac{\Lambda_H^2}{2 M_2^2 \Delta} \biggr) \nonumber\\
&\times& \exp\Bigg \{-\frac{1}{\Lambda_H^2} \biggl[ R(\alpha,\beta)
\, M_H^2 \,+\,\alpha \, M_1^2\,+\,\beta \, M_2^2 \Big ]\Bigg \}\,.
\en

\begin{figure}
\centering{ \includegraphics[scale=0.55]{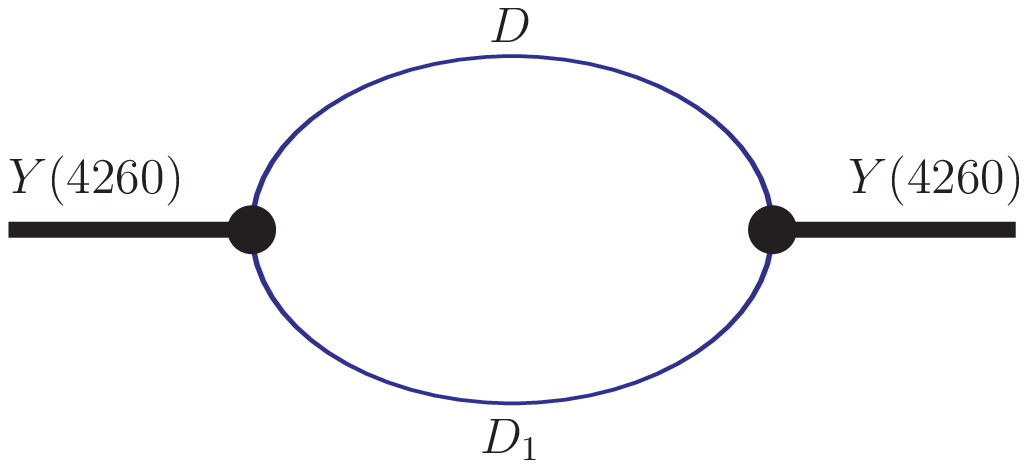}
\hspace*{1cm}
\includegraphics[scale=0.55]{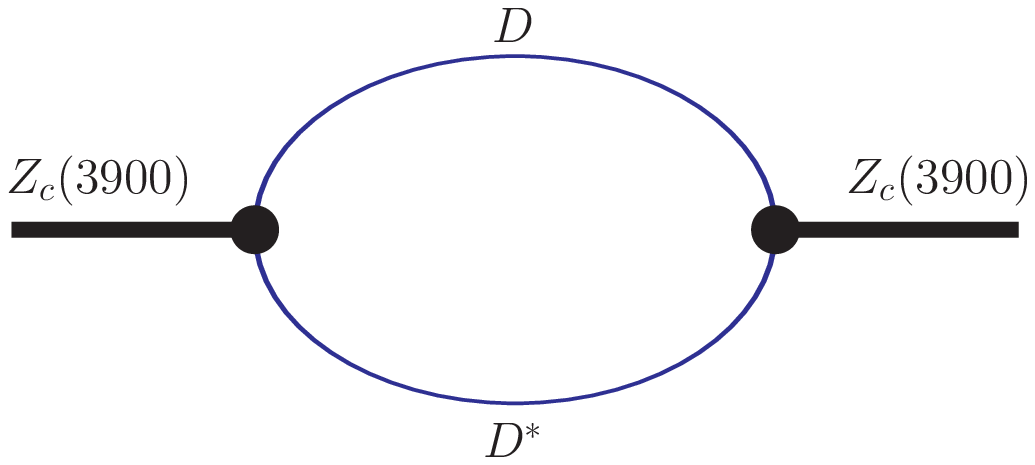}}
\caption{Mass operators of $Z_c(3900)$ and $Y(4260)$.}

\vspace*{1cm}

\centering \includegraphics[scale=0.75]{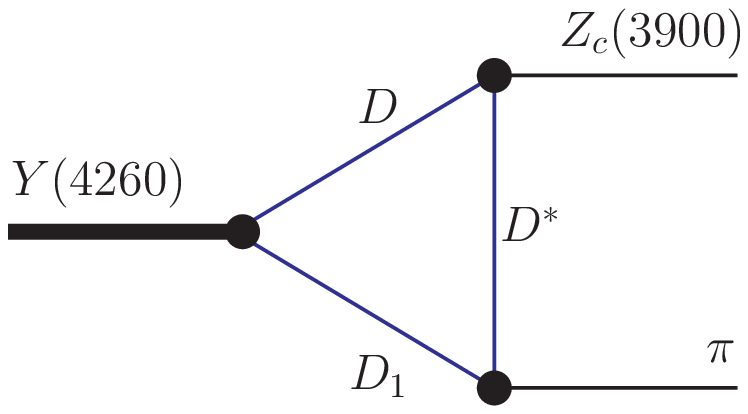}
\caption{Diagram contributing to the
$Y(4260) \to Z_c(3900) + \pi$ decay.}

\vspace*{1cm}

\centering \includegraphics[scale=0.75]{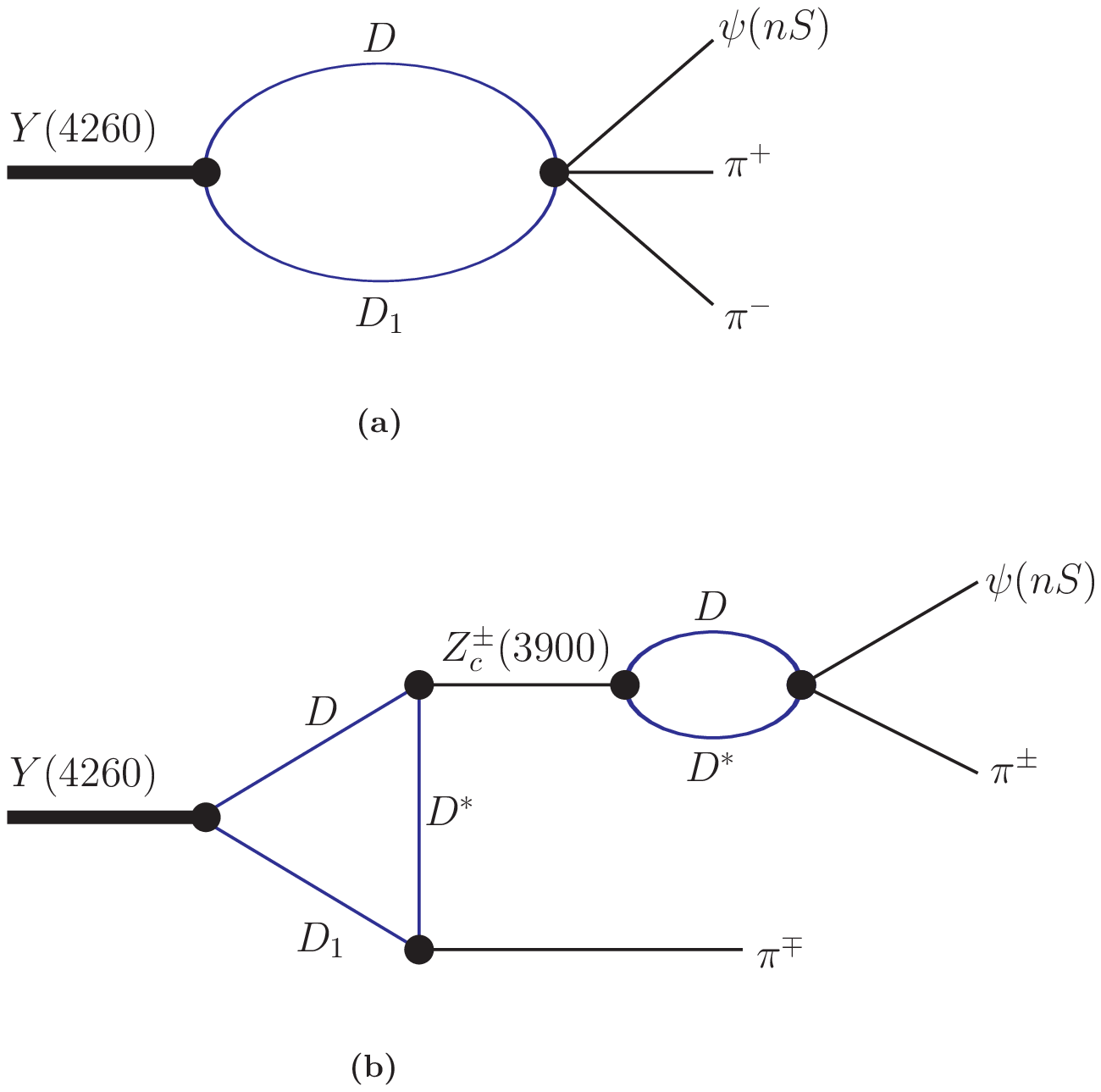}
\hspace*{.5cm}
\caption{Diagrams contributing to the
$Y(4260) \to \psi(nS)+\pi^{\pm}$ decay: (a) contact diagram,
(b) $Z_c(3900)$ resonance diagrams.}
\end{figure}

In the previous equation we use the notation
\eq
\Delta=2+\alpha+\beta, \quad R(\alpha,\beta) =
\frac{(2\omega+\alpha)^2}{\Delta} - \biggl( 2\omega^2 + \alpha
\biggr)\,,
\en
where
$\omega=\frac{M_2}{M_1+M_2}$, $(M_1,M_2)= (M_D, M_{D^*})$
for $H=Z_c$ and $(M_D, M_{D_1})$ for $H=Y$, respectively.

In the calculation we use for the $Y(4260)$, the mass $M_Y =
4250\pm 9$ MeV and width
$\Gamma_Y = 108\pm 12$ MeV~\cite{Coan:2006rv}.
The mass of the $Z_c(3900)$ is expressed in terms of the constituent
meson masses and the binding energy $\epsilon_{Z_c}$ as
\eq
M_{Z_c}=M_{D^0}+M_{D^{\ast\,0}}-\epsilon_{Z_c}\,.
\en
Note that $\epsilon_{Z_c}$ is a variable quantity in our calculations,
which we vary
from 0.5 to 5 MeV.  Once the mass of the composite state ${Z_c}$ is fixed,
the value for the coupling of $Z_c(3900)$ to $DD^*$ can be extracted from
the compositeness condition. Values for this coupling in dependence on the
binding energy $\epsilon_{Z_c}$ and the cutoff $\Lambda_{Z_c}$
are shown in Table I. Note that the dependence of $g_{Z_c}$ on the binding
energy is in agreement with the scaling law of
hadronic molecules found in Ref.~\cite{Branz:2008cb}:
$g_{Z_c} \sim \epsilon_{Z_c}^{1/4}$. Moreover, for the central
mass value of the Y(4260), the couplings of Y(4260) to $DD_1$ are 7.85 for
$\Lambda_Y=0.5$ GeV and 6.25 for $\Lambda_Y=0.75$ GeV, respectively.

\begin{center}

{\bf Table I.} Results for the coupling constant $g_{Z_c}$
depending on $\epsilon_{Z_c}$ and $\Lambda_{Z_c}$.

\vspace*{.25cm}
\begin{tabular}{|c|c|c|c|c|}\hline
                   & \multicolumn{4}{c|}{}  \\
$\Lambda_{Z_c}$ in GeV
&\multicolumn{4}{c|}{$\epsilon_{Z_c}$ in MeV}\\[1mm]
\cline{2-5}
     &0.5 &1   &2.5 &5   \\ \hline
0.5  &2.2 &2.3 &2.6 &3.1 \\ \hline
0.75 &2.1 &2.2 &2.5 &2.9 \\ \hline
\end{tabular}
\end{center}

\vspace*{.25cm}

\begin{center}
{\bf Table II.} Decay widths for $Y(4260) \to Z_c(3900)^+ + \pi^-$
in MeV.

\vspace*{.1cm}
\def\arraystretch{1.5}
\begin{tabular}{|c|c|c|c|c|}\hline
                   & \multicolumn{4}{c|}{}  \\
$\e_{Z_c}$ in MeV
&\multicolumn{4}{c|}{$(\Lambda_{Z_c}, \Lambda_Y)$ in GeV}\\[1mm]
\cline{2-5}
& (0.5,0.5) & (0.5,0.75) & (0.75,0.5) &(0.75,0.75)  \\
\hline 0.5     &$3.1$  &$2.7$  &$3.1$  &$2.9$  \\
\hline 1       &$3.1$  &$2.7$  &$3.2$  &$3.0$  \\
\hline 2.5     &$3.3$  &$2.9$  &$3.4$  &$3.2$  \\
\hline 5       &$3.4$  &$3.1$  &$3.7$  &$3.6$  \\
\hline
\end{tabular}
\end{center}

The diagram contributing to the two-body decays $Y(4260)\to Z_c(3900) + \pi$
is shown in Fig.~2. The diagrams contributing to the
$Y(4260)\to \psi(nS) + \pi^+\pi^-$ transition are drawn in Fig.~3:
the contact diagram [Fig.~3(a)] and the resonance diagram [Fig.~3(b)].
For the $Y(4260)$ decays, as presented by the diagrams of Figs.~2 and 3,
additional dynamical input is needed.
To calculate the two-body decays $Y(4260)\to Z_c(3900) + \pi$
(Fig.~2) and the resonance diagram corresponding to the
transition $Y(4260)\to Z_c^{\pm}(3900)+\pi^{\mp}\to
J/\psi \pi^+\pi^-$ (see Fig.~3(b)), we use
the phenomenological Lagrangian for the $D_1\to D^\ast\pi$ coupling with
\eq
{\cal L}_{D_1 D^\ast\pi}=\frac{g_{_{D_1}}}{2\sqrt{2}}\bar{D}_1^{\mu\nu}\,
\vec{\pi} \vec{\tau} \,
D^\ast_{\mu\nu} \,+\, {\rm H.c.}\,,
\en
where $D_1^{\mu\nu}$ and $D^\ast_{\mu\nu}$ are the stress tensors
of $D_1$ and $D^\ast$ mesons.
The coupling $g_{_{D_1}}$ can be estimated by considering the decay width
of $D_1(2420)\to D^{\ast\,+}+\pi^-$ of  $\sim 20$~MeV (see details
in Ref.~\cite{Branz:2010sh})
which is 2/3 of the total $D_1(2420)$ width. Then one gets
$g_{_{D_1}} \simeq 0.49$~GeV$^{-1}$.

For the evaluation of the contact diagram in Fig.~3(a),
we also need the $DD_1\pi^+\pi^-\psi$ interaction Lagrangian which
can be derived using formalism proposed in Ref.~\cite{Dong:2013iqa}.
In particular, one can derive a phenomenological Lagrangian
describing the coupling of heavy quarkonia with pair of pseudoscalar,
vector or axial heavy-light mesons and light pseudoscalar mesons.
The phenomenological Lagrangian reads
\eq\label{Lagr_eff}
{\cal L}_{\cal D \bar D H P}(x) =
i g_F \,{\rm tr} \Big( \bar{\cal D}(x) \,
[ {\cal H}(x), {\cal P}(x)] \, {\cal D}(x) \Big)
\, + \,
g_D \,{\rm tr} \Big( \bar{\cal D}(x) \,
\{ {\cal H}(x), {\cal P}(x)\} \, {\cal D}(x) \Big)
\,,
\en
where $g_F$ and $g_D$ are effective coupling constants,
$[\ldots]$ and $\{\ldots\}$ denote the commutator and
anticommutator, respectively.

The $H$ is the heavy charmonia field; $D$ is the superposition of
isodoublets of open-charm mesons with $J^P = 0^-, 1^-$ and $1^+$;
${\cal P}$ is the chiral field:
\eq
{\cal H} &=& J^\mu \gamma_\mu + h^\mu \gamma_\mu \gamma_5
\, + \, \frac{g_{H}}{M_H} \,
\Big( J^{\mu\nu} \, \sigma_{\mu\nu} \, +  \,
h^{\mu\nu} \sigma_{\mu\nu} \gamma_5 \Big) \,, \\
{\cal D} &=& D i\gamma_5 + D^{\ast\,\mu} \gamma_\mu
          +  D_1^{\mu} \gamma_\mu \gamma_5  \,,\\
{\cal P} &=& \frac{1}{2} \not\! u \, \gamma^5
          + \frac{1}{2} [u^\dagger,\partial_\mu u] \gamma^\mu
          + g_p u_\mu u^\mu \,,
\en
where $J$ and $h$ denote the $\Psi$ and $h_c$ states;
$V^{\mu\nu} = \partial^\mu V^\nu - \partial^\nu V^\mu$ is
the stress tensor of the $\Psi$ and $h_c$ states;
$g_{\cal H}$ is a  phenomenological
coupling  defining the mixing of derivative and nonderivative terms
in ${\cal H}$; $M_H \simeq M_J$ is associated with the $J/\psi$ mass;
$D = (D^+, D^0)$, $D^\ast = (D^{\ast\, +}, D^{\ast\, 0})$
are the doublets of pseudoscalar and vector
charmed $D$ mesons; $u_\mu$ is the chiral vielbein:
\eq
u_\mu = i \{ u^\dagger, \partial_\mu u \}\,, \quad u^2 = U
= \exp\Big[i\frac{\hat{\pi}}{F_\pi}\Big]\,, \quad
\hat{\pi} = \vec{\pi} \vec{\tau}
\en
where $F_\pi = 92.4$ MeV is the pion decay constant,
$\vec{\pi}=(\pi_1,\pi_2,\pi_3)$ is the triplet of pions.
Note the couplings $g_D$, $g_F$ and $g_H$ are phenomenological
parameters. Below we show that we have two constraints on these couplings.

From Eq.~(\ref{Lagr_eff}) we deduce specific Lagrangians describing the
couplings between heavy charmonia, charmed mesons and the pion which
are relevant for the decay $Y(4260) \to J/\psi + \pi^+ \pi^-$
\eq
{\cal L}_{\pi\pi DD_1\psi}=\frac{g_{_{DD_1}}}{F_{\pi}^2} \,
J^{\mu} \, \bar{D}_{1\mu}D \,
\partial^\nu\vec{\pi} \, \partial_\nu\vec{\pi} \,+\, {\rm H.c.}\,,
\quad g_{_{DD_1}} = 4 g_p g_D =0.86 GeV^{-1} 
\en
In the evaluation of Fig.~3(b) the subprocess $Z_c^\pm \to (D
D^\ast)^\pm \to \psi(nS) + \pi^\pm$ is treated as worked out in
Ref.~\cite{Dong:2013iqa}.

\section{Decay modes and results}

The two-body decay width for the transition $Y(4260)\to Z_c(3900) +
\pi$ described in Fig.~2 is given by
\eq
\Gamma_2 \, = \, \int\limits_{M_{_{Z_c}}+M_{\pi}}^{M_Y+3\Gamma_Y}\,
dm_Y \,
f(m_Y) \, \frac{|{\bf p}(m_Y)|}{24\pi \, m_Y^2} \
\sum_{\rm pol} \Big|M_{\rm inv,2}(M_Y)\Big|^2
\en
where $|{\bf p}(m_Y)| = \lambda^{1/2}(m_Y^2,M_{Z_c}^2,M_\pi^2)/(2m_Y)$
is the magnitude of the three-momentum of outgoing particles in
the rest frame of the $Y(4260)$ state,
\eq
\lambda(x,y,z)=x^2+y^2+z^2-2xy-2yz-2xz
\en
is the K\"allen function and $\M_{\rm inv,2}$ is the
corresponding invariant matrix element. We consider the finite width of
the Y(4260) by setting up a mass distribution $f(m_Y)$ in the
form~\cite{Dover:1990kn,Gutsche:2008qq}
\eq
f(m_Y) = \left\{
\begin{array}{ll}
0\,, & \hspace*{1cm} m_Y < M_{\rm thr} \\
\displaystyle\frac{1}{4 A_0} \cdot \frac{\Gamma_Y^2}{(m_Y-M_Y)^2 +
\frac{1}{4} \Gamma_Y^2}
\cdot \frac{m_Y-M_{\rm thr}}{M_Y-\Gamma_Y-M_{\rm thr}} \,,
& \hspace*{1cm}  M_{\rm thr} < m_Y < M_Y-\Gamma_Y \\
\displaystyle\frac{1}{4 A_0} \cdot
\frac{\Gamma_Y^2}{(m_Y-M_Y)^2+\frac{1}{4} \Gamma_Y^2} \,,
& \hspace*{1cm} M_Y - \Gamma_Y < m_Y < M_Y + \Gamma_Y \\
\displaystyle\frac{1}{4 A_0} \cdot \frac{\Gamma_Y^2}{(m_Y-M_Y)^2
+ \displaystyle\frac{1}{4} \Gamma_Y^2} \cdot
\frac{M_Y+3\Gamma_Y - m_Y}{2 \Gamma_Y}\,,
& \hspace*{1cm}  M_Y +  \Gamma_Y < m_Y < M_Y + 3 \Gamma_Y \\
0 \,, & \hspace*{1cm} M_Y + 3 \Gamma_Y < m_Y \,.
\end{array} \right.
\en
The lowest strong decay threshold is denoted by
$M_{\rm thr} = M_{J/\psi} + 2M_{\pi} \, \simeq \, 3.376$ GeV, and $A_0$
is a normalization constant such that
\eq
\int\limits_0^\infty \, dm_Y \, f(m_Y) \, = \, 1 \,.
\en
In Eq. (13) we average the available phase space over the mass distribution
of the $Y(4260)$, while the matrix element is evaluated at the central mass
value of $4250$ MeV. This procedure will capture the major features of
the mass distribution of the $Y(4260)$.

The three-body decay width related to the process of Fig.~3 is
evaluated as
\eq
\Gamma_3 \, = \, \int\limits_{\psi(nS)+2M_{\pi}}^{M_Y+3\Gamma_Y} \,
dm_Y \, \frac{f(m_Y)}{768\pi^3m_Y^3} \,
\int\limits^{(m_Y-m_1)^2}_{(m_2+m_3)^2}ds_2
\int\limits^{s_1^+}_{s_1^-}ds_1 \, \sum_{\rm pol}
\Big|M_{\rm inv,3}(M_Y)\Big|^2
\en
where $p_{1,2,3}$ and $m_{1,2,3}$ are the momenta and masses of the
three outgoing particles; $s_1 = (p_1 + p_2)^2$ and $s_2 = (p_2 +
p_3)^2$ are the Mandelstam variables,
\eq
s_1^{\pm}=m_1^2+m_2^2-\frac{1}{2s_2} \Big
((s_2-m_Y^2+m_1^2)(s_2+m_2^2-m_3^2)\mp\lambda^{1/2}(s_2,m_Y^2,m_1^2)
\lambda^{1/2}(s_2,m_2^2,m_3^2)\Big )\,.
\en
Again, the invariant matrix element of the three-body
decay is simply denoted by $\M_{\rm inv,3}(M_Y)$, which is estimated
at the central $Y(4260)$ mass $M_Y=4250$ MeV. The calculation of phase space
includes the mass distribution of the $Y(4260)$. The evaluation of the
invariant matrix elements in both the two- and three-body decay is
standard and not explicitly written out. The calculational technique
is, for example, discussed in detail in Ref.~\cite{Dong:2013iqa}.

The diagram contributing to the two-body decay $Y(4260) \to Z_c(3900) + \pi$
is shown in Fig.~2. Numerical results for $\Gamma(Y(4260) \to Z_c(3900)^+
+ \pi^-)$, which are of the order of a few MeV, are given both in Table II and
Fig.~4. In our calculations we use two different values for the cutoff
parameters $\Lambda_{Y}$ and $\Lambda_{Z_c}$ --- $0.5$ and $0.75$~GeV.
The dependence of the decay width on the size parameter is only moderate.
A larger binding energy $\epsilon_{Z_c}$ leads to an increase in phase space
and hence in the decay width.

\vspace*{.25cm}
\begin{center}
{\bf Table III.} $Y(4260) \to J/\psi(\psi(2S)) + \pi^+ \pi^-$ decay
widths in MeV. \\ Predictions for the mode with $\psi(2S)$ are given
in brackets. \\
For total width of $Z_c(3900)$, we use BESIII data
$\Gamma_{Z_c} = 46\pm 10\pm 20$ MeV~\cite{Ablikim:2013mio}.

\vspace*{.25cm}
\def\arraystretch{1.5}
\begin{tabular}{|c|c|c|c|c|}\hline
                   & \multicolumn{4}{c|}{}  \\
$\e_{Z_c}$ in MeV
&\multicolumn{4}{c|}{$(\Lambda_{Z_c}, \Lambda_Y)$ in GeV}\\[1mm]
\cline{2-5}
& (0.5,0.5) & (0.5,0.75) & (0.75,0.5) &(0.75,0.75)\\
\hline
0.5     &$0.3^{-0.01}_{+0.01}$      &$0.8^{-0.01}_{+0.01}$
        &$0.4^{-0.02}_{+0.07}$   &$0.8^{-0.01}_{+0.03}$ \\
        &($0.2^{-0.01}_{+0.04}$) &($0.4^{-0.02}_{+0.05}$)
        &($0.2^{-0.03}_{+0.09}$) &($0.5^{-0.04}_{+0.10}$) \\
\hline
1       &$0.3^{-0.01}_{+0.02}$   &$0.8^{-0.01}_{+0.01}$
        &$0.4^{-0.02}_{+0.09}$   &$0.8^{-0.01}_{+0.06}$  \\
        &($0.2^{-0.02}_{+0.04}$) &($0.4^{-0.02}_{+0.05}$)
        &($0.2^{-0.04}_{+0.1}$)  &($0.5^{-0.04}_{+0.1}$) \\
\hline
2.5     &$0.4^{-0.01}_{+0.6}$   &$0.8^{-0.01}_{+0.04}$
        &$0.4^{-0.04}_{+0.2}$    &$0.8^{-0.03}_{+0.1}$\\
        &($0.2^{-0.02}_{+0.07}$) &($0.4^{-0.03}_{+0.07}$)
        &($0.3^{-0.05}_{+0.1}$)  &($0.5^{-0.06}_{+0.2}$)   \\
\hline
5       &$0.4^{-0.06}_{+0.2}$   &$0.9^{-0.04}_{+0.2}$
        &$0.5^{-0.1}_{+0.4}$   &$0.9^{-0.08}_{+0.4}$   \\
        &($0.3^{-0.05}_{+0.1}$) &($0.5^{-0.05}_{+0.2}$)
        &($0.3^{-0.08}_{+0.3}$) &($0.6^{-0.08}_{+0.2}$) \\
\hline
\end{tabular}
\end{center}

\begin{center}
{\bf Table IV.} $Y(4260) \to J/\psi(\psi(2S)) + \pi^+ \pi^-$ decay
widths in MeV. \\ Predictions for the mode with $\psi(2S)$ are given
in brackets. \\
For total width of $Z_c(3900)$, we use Belle data
($63\pm 24\pm 26$ MeV)~\cite{Liu:2013dau}.

\vspace*{.25cm}
\def\arraystretch{1.5}
\begin{tabular}{|c|c|c|c|c|}\hline
                   & \multicolumn{4}{c|}{}  \\
$\e_{Z_c}$ in MeV
&\multicolumn{4}{c|}{$(\Lambda_{Z_c}, \Lambda_Y)$ in GeV}\\[1mm]
\cline{2-5}
& (0.5,0.5) & (0.5,0.75) & (0.75,0.5) &(0.75,0.75)\\
\hline
0.5     &$0.3^{-0.01}_{+0.01}$   &$0.8^{-0.01}_{+0.01}$
        &$0.3^{-0.01}_{+0.06}$   &$0.8^{-0.01}_{+0.02}$ \\
        &($0.3^{-0.01}_{+0.04}$)  &($0.4^{-0.01}_{+0.05}$)
        &($0.2^{-0.02}_{+0.09}$) &($0.4^{-0.03}_{+0.1}$) \\
\hline
1       &$0.3^{-0.01}_{+0.02}$   &$0.8^{-0.01}_{+0.01}$
        &$0.4^{-0.01}_{+0.08}$   &$0.8^{-0.01}_{+0.04}$  \\
        &($0.3^{-0.01}_{+0.04}$) &($0.4^{-0.02}_{+0.05}$)
        &($0.2^{-0.03}_{+0.1}$)  &($0.4^{-0.03}_{+0.1}$) \\
\hline
2.5     &$0.3^{-0.01}_{+0.05}$   &$0.8^{-0.01}_{+0.02}$
        &$0.4^{-0.02}_{+0.2}$     &$0.8^{-0.01}_{+0.1}$\\
        &($0.2^{-0.02}_{+0.06}$)  &($0.4^{-0.02}_{+0.07}$)
        &($0.2^{-0.03}_{+0.1}$)   &($0.2^{-0.04}_{+0.2}$)   \\
\hline
5       &$0.4^{-0.03}_{+0.2}$   &$0.8^{-0.02}_{+0.2}$
        &$0.5^{-0.06}_{+0.4}$   &$0.9^{-0.04}_{+0.3}$   \\
        &($0.2^{-0.03}_{+0.1}$) &($0.5^{-0.04}_{+0.2}$)
        &($0.3^{-0.05}_{+0.2}$) &($0.5^{-0.06}_{+0.2}$) \\
\hline
\end{tabular}
\end{center}

\vspace*{.25cm}

\begin{center}
{\bf Table V.} Contribution of the contact diagram in Fig.~3(a) \\
to $\Gamma(Y(4260) \to J/\psi(\psi(2S)) + \pi^+ \pi^-)$ in MeV. \\

\vspace*{.1cm}
\def\arraystretch{1.5}
\begin{tabular}{|c|c|c|}\hline
Mode    & $\Lambda_Y = 0.5$  GeV
        & $\Lambda_Y = 0.75$ GeV \\
\hline
$Y(4260) \to J/\psi + \pi^+ \pi^-$
& $0.26$
& $0.64$ \\
\hline
$Y(4260) \to \psi(2S) + \pi^+ \pi^-$
& $0.12$
& $0.30$ \\
\hline
\end{tabular}
\end{center}

In Tables III and IV we present our numerical results for the
widths of both decay modes involving $J/\psi$ and $\psi(2S)$ (in brackets).
Results are given for different values of the $Z_c$ binding energy
and of the respective size parameters $\Lambda_Y$ and $\Lambda_{Z_c}$.
In the calculation of the $Z_c$ resonance contribution,
the propagator of $Z_c$ state is described by a Breit--Wigner form,
where we have used a constant width $\Gamma_{Z_c}$ in the imaginary part,
i.e. we have used
\begin{equation}
D_{Z_c}(q^2) = \frac{1}{M^2_{Z_c}-q^2-i M_{Z_c}\Gamma_{Z_c}} \,.
\end{equation}
We consider two results for
$\Gamma_{Z_c}$ including error bars:
$46\pm 10\pm 20$ MeV --- according to  BESIII~\cite{Ablikim:2013mio}
(see Table III) and --- $63\pm 24\pm 26$ MeV --- the result of the
Belle Collaboration~\cite{Liu:2013dau} (see Table IV).
The results for the sole contribution of the contact
diagram in Fig.~3(a) are shown in Table V.

For completeness we also plot our results for the three-body decays in
Figs.~5-7. In particular, in Fig.~5 we plot the ratio $R$ of resonance
diagram [Fig.~3(b)] to total contribution (Fig.~3) to the decay width
$Y(4260) \to J/\psi + \pi^+ \pi^-$ as a function of the binding energy
$\epsilon_{Z_c} = 0.5 - 5$ MeV and for different cutoff parameters
$\Lambda_Y = 0.5, 0.75$ GeV. The two horizontal lines at $R = 0.201$
and $R = 0.378$ define the lower and upper limit set by data of the Belle
Collaboration~\cite{Liu:2013dau},
\eq
\label{ratio_R}
R = \frac{{\rm Br}(Y(4260)\to Z_c^{\pm}\pi^{\mp}) \,
{\rm Br}(Z_c^{\pm} \to J/\psi \pi^{\pm})}
{{\rm Br}(Y(4260)\to J/\psi\pi^+\pi^-)} \, = \, (29.0\pm 8.9)\%.
\en
As is evident from Fig.~5,
the ratio $R$ is rather sensitive to explicit values of
the binding energy $\epsilon_{Z_c}$ and the choice of size parameters. The
present range of values set by Belle can be reproduced in the calculation for
restricted values of the varied quantities.
In Figs.~6 and 7 we plot the total contributions to the decay
rates of $Y(4260)\to J/\psi\pi^+\pi^-$ and $Y(4260)\to \psi(2S) \pi^+\pi^-$.
In Fig.~6 we indicate the horizontal line $\Gamma(Y(4260)\to J/\psi\pi^+\pi^-)
= 0.508$ MeV corresponding to the lower limit for this decay rate extracted
from data of Ref.~\cite{Mo:2006ss}. From this constraint and from data for the
ratio $R$ we estimate that the lower limit of $\Gamma(Y(4260)\to
Z_c^{\pm}\pi^{\mp}\to J/\psi\pi^+\pi^-)$ is larger than 100 keV.
Using the last constraint and the results for the ratio $R$,
we conclude that for
a favored value of $\Lambda_Y= 0.75$ GeV we deduce a lower limit
of $\Gamma(Y(4260)\to J/\psi\pi^+\pi^-)>0.8$ MeV.
Our results for the favored value  of $\Lambda_Y= 0.75$ GeV
and including the variation of the cutoff parameter $\Lambda_{Z_c}$
and of the binding energy $\epsilon_{Z_c}$ can be summarized as
\eq
\Gamma(Y(4260)\to Z_c^{\pm}(3900)+\pi^{\mp}) &=& 3.15 \pm 0.45 \
\mathrm{MeV}\,,
\nonumber\\
\Gamma(Y(4260) \to J/\psi + \pi^+ \pi^-)  &=& 1 \pm 0.20 \
\mathrm{MeV}\,, \\
\Gamma(Y(4260) \to \psi(2S) + \pi^+ \pi^-) &=& 0.55 \pm 0.15 \
\mathrm{MeV}
\nonumber\,.
\en

In summary, using a phenomenological Lagrangian approach,
we give predictions for the two- and three-body decay rates
$Y(4260)\to Z_c^{\pm}(3900)+\pi^{\mp}$ and $Y(4260)\to \psi(nS) + \pi^+\pi^-$
for $n=1,2$. Our results for the two-body decays are in the order
of several MeV. For the $Y(4260)$ three-body decay with a charged pion pair,
we estimated both contact and $Z_c(3900)$-resonance
contributions, and the decay rate varies from several hundred keV
to a few MeV. Both the background and $Z_c(3900)$ contributions are
explicitly shown. We expect that the quantitative predictions given here
can serve as a further test for the molecular interpretation of the
$Y(4260)$ and can be measured in forthcoming experiments.

In further work we plan to estimate other strong and also radiative
decay modes of the $Y(4260)$ state. In particular,
the following decay modes of the  $Y(4260)$ state can be analyzed:
strong decay modes with two heavy charm mesons $DD$,
$DD^\ast$ or $D^\ast D^\ast$ in the final state, the strong decay modes with
a $h_c$ state, and radiative decay $Y(4260) \to X(3872) + \gamma$.
Special attention will be paid to an analysis of other possible hidden
charm resonances [in addition to $Z_c(3900)$] with spin-parities
$0^\pm, 1^\pm$ and with a mass in the interval $3900 - 4100$ MeV,
all contributing to the total width of the $Y(4260)$ state.

\begin{acknowledgments}

We thank Alex Bondar, Qiang Zhao, Dian Yong Chen and Changzheng Yuan
for useful discussions.
This work is supported by the DFG under Contract No. LY 114/2-1,
the National Sciences Foundations of China No.10975146 and No.11035006,
and by the DFG and the NSFC through funds provided to the sino-German CRC 110
``Symmetries and the Emergence of Structure in QCD.''
The work is done partially under the Project No. 2.3684.2011 of Tomsk State
University. V. E. L. would like to thank Tomsk Polytechnic University, Russia
for warm hospitality.
Y. B. D. thanks the Institute of Theoretical Physics,
University of T\"ubingen, for warm hospitality and
the Alexander von Humboldt Foundation for support.

\end{acknowledgments}

\newpage

\begin{figure}
\centering \includegraphics[scale=0.5]{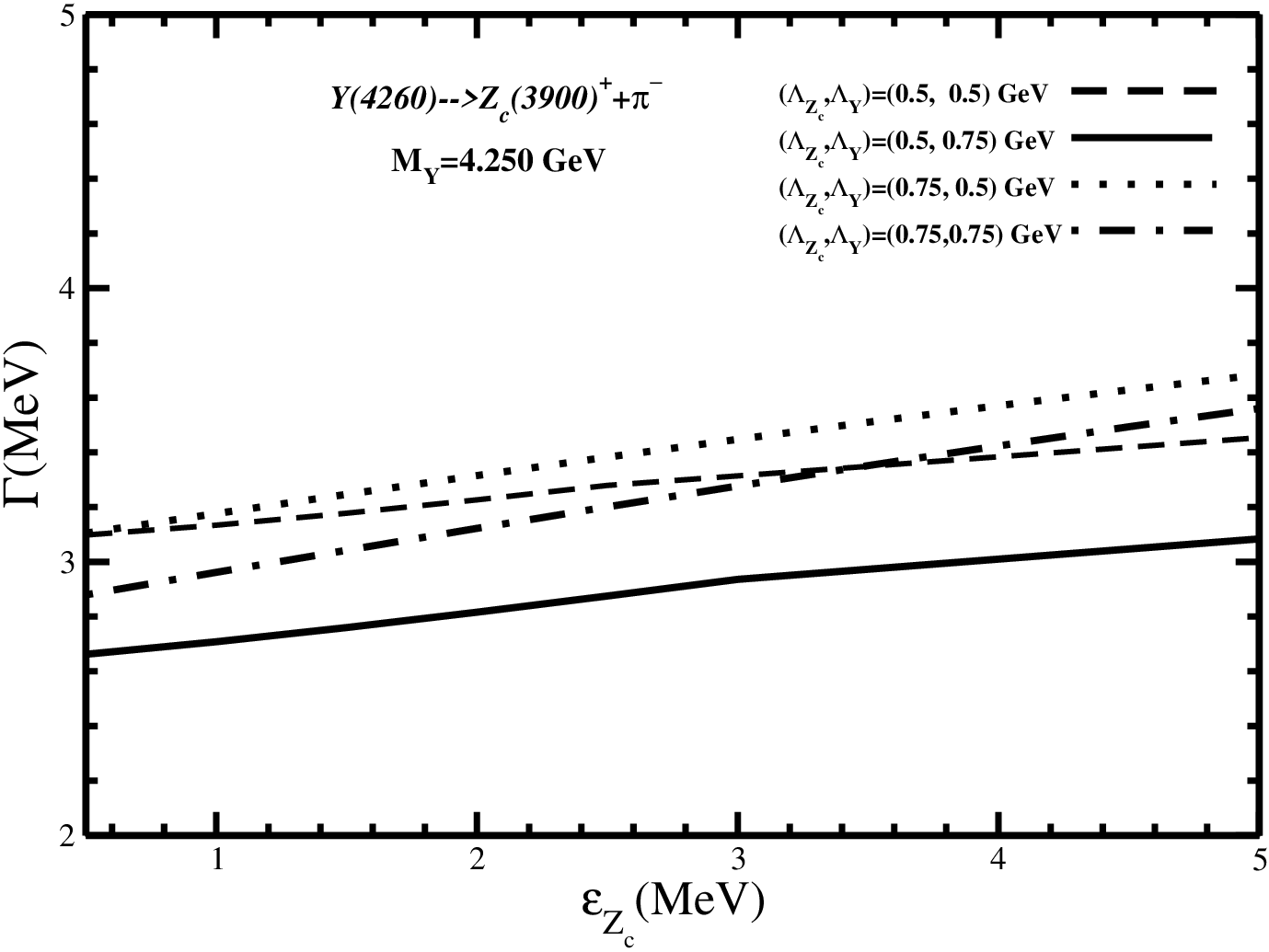}
\caption{Decay width $Y(4260) \to Z_c(3900)^+ + \pi^-$
as function of binding energy
$\epsilon_{Z_c} = 0.5 - 5$ MeV
and for different cutoff parameters $\Lambda_Y = 0.5, 0.75$ GeV.}

\vspace*{1.5cm}

\centering \includegraphics[scale=0.5]{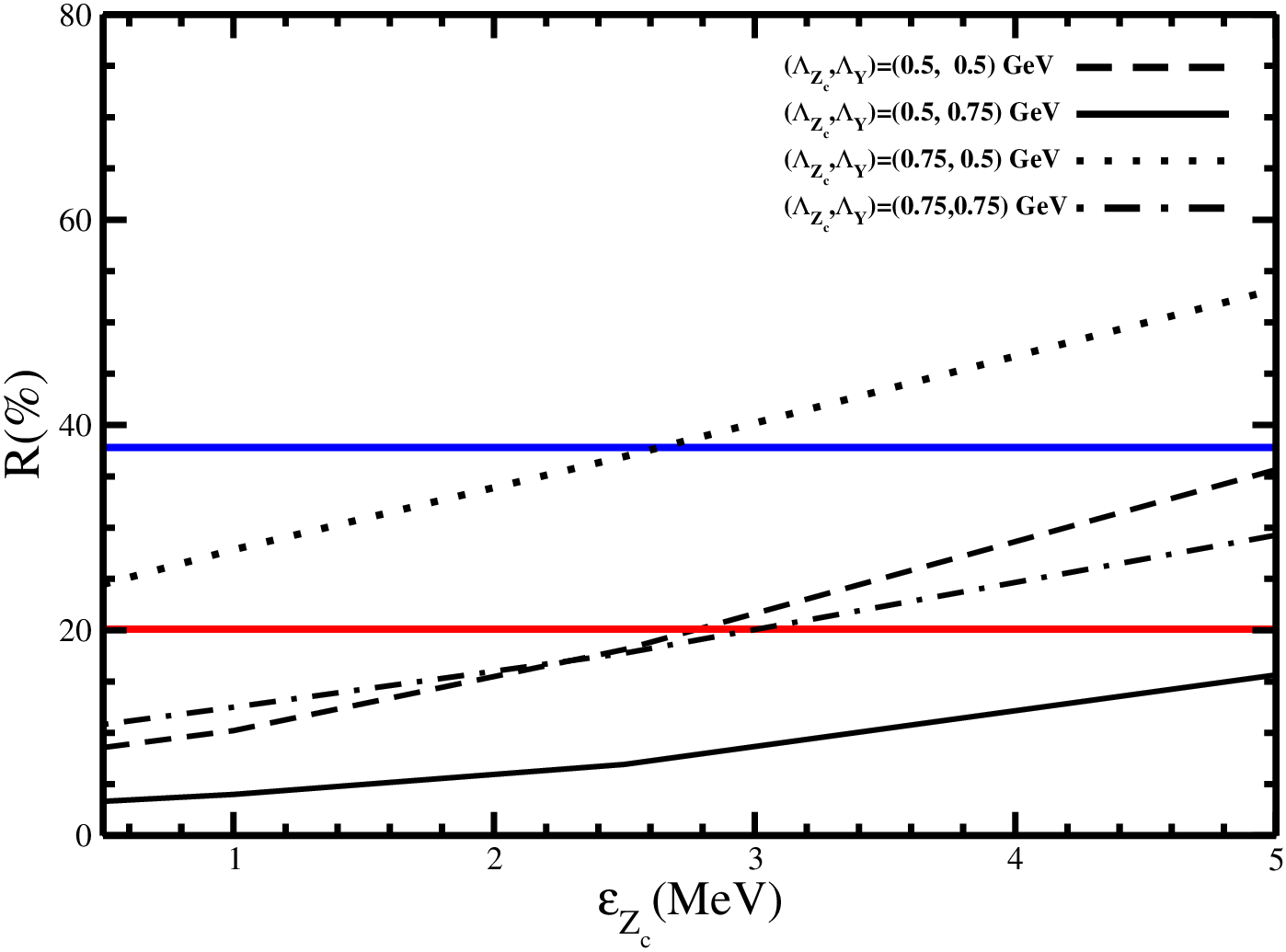}
\caption{Ratio $R$ of the resonance diagram contribution in Fig.~3(b)
to the total one (Fig.~3) for the decay width
$Y(4260) \to J/\psi + \pi^+ \pi^-$ as function
of the binding energy $\epsilon_{Z_c} = 0.5 - 5$ MeV
and for different cutoff parameters $\Lambda_Y = 0.5, 0.75$ GeV.
The two horizontal lines at $R = 0.201$ and $R = 0.378$ define
the lower and upper limit of data from the Belle
Collaboration~\cite{Liu:2013dau}.}

\end{figure}

\newpage

\begin{figure}

\centering \includegraphics[scale=0.5]{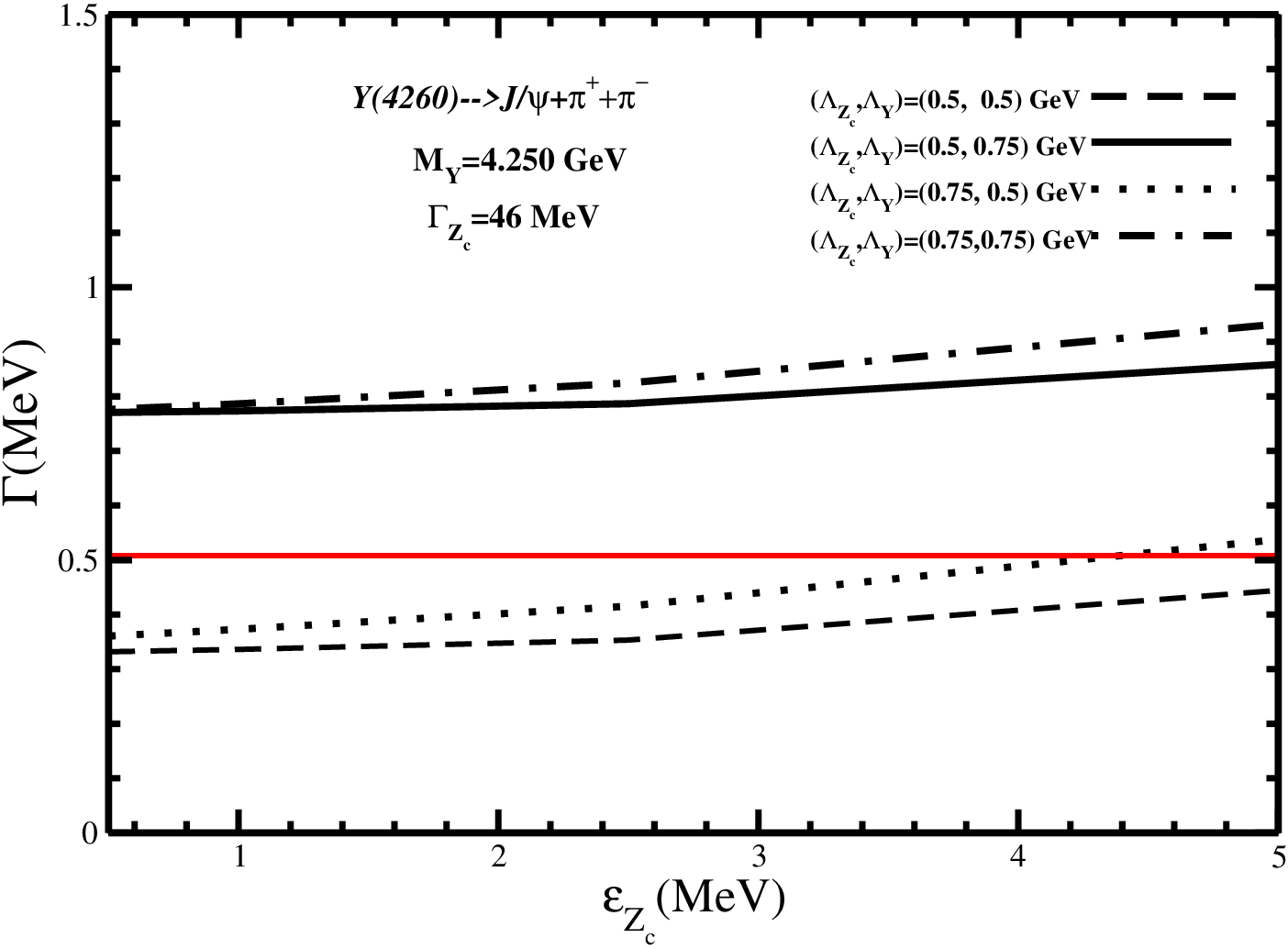}
\caption{Total contribution to the decay width
$Y(4260) \to J/\psi + \pi^+ \pi^-$
as function of the binding energy $\epsilon_{Z_c} = 0.5 - 5$ MeV
and for different values of the cutoff parameters
$\Lambda_{Z_c}$ and $\Lambda_Y = 0.5, 0.75$ GeV.
The horizontal line $\Gamma(Y(4260) \to J/\psi + \pi^+ \pi^-)
= 0.508$~GeV corresponds to the lower limit deduced from data
in Ref.~\cite{Mo:2006ss}.}

\vspace*{1.5cm}

\centering \includegraphics[scale=0.5]{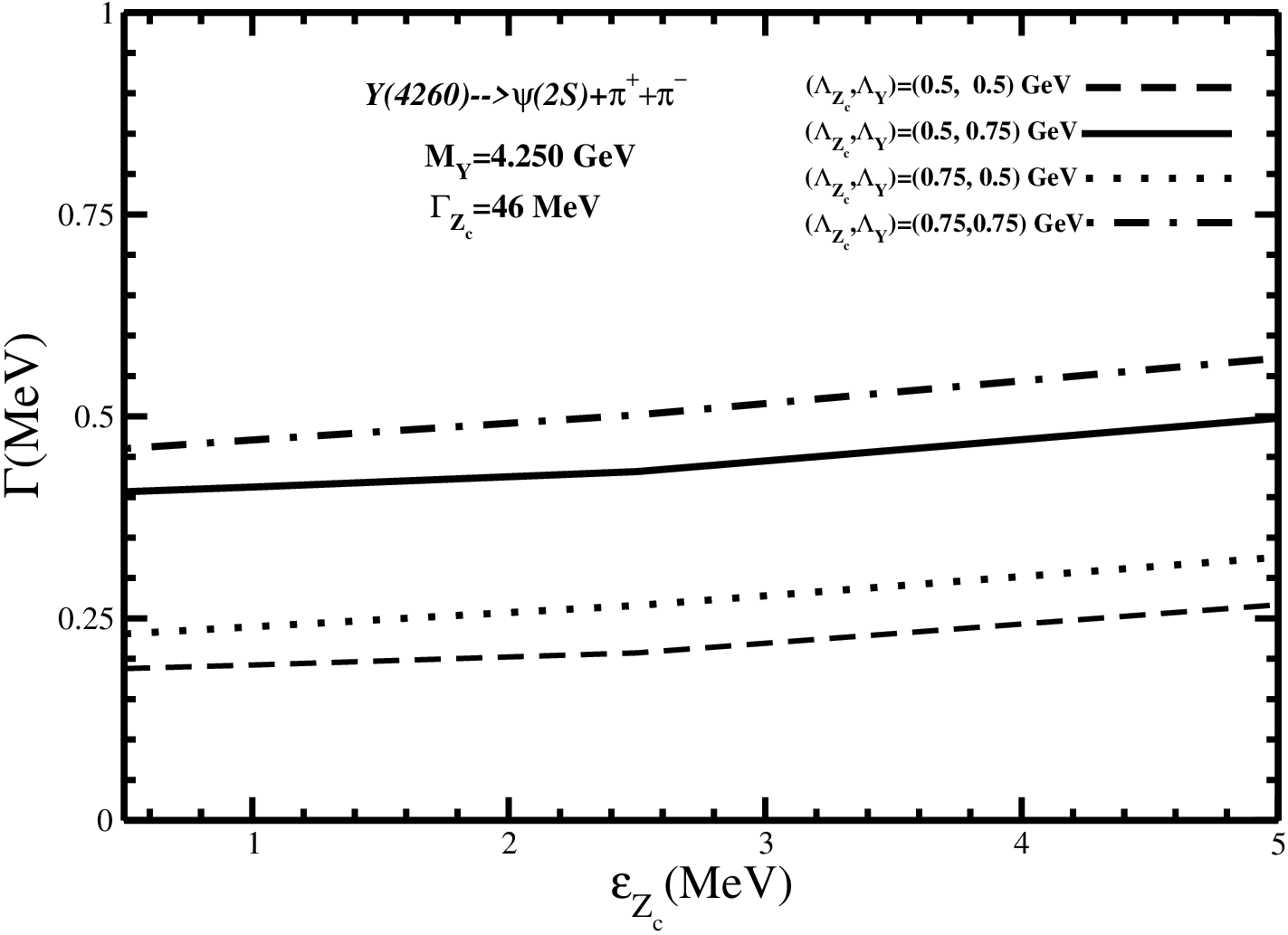}
\caption{Total contribution to the decay width
$Y(4260) \to \psi(2S) + \pi^+ \pi^-$
as function of the binding energy $\epsilon_{Z_c} = 0.5 - 5$ MeV
and for different values of the cutoff parameters
$\Lambda_{Z_c}$ and $\Lambda_Y = 0.5, 0.75$ GeV.}

\end{figure}

\end{document}